\documentclass[12pt,preprint]{aastex}
\usepackage{graphics,graphicx}
\usepackage{fancyheadings}
\usepackage{color}
\usepackage{natbib}

\newcommand{\pks}{PKS\,1510$-$089}

\shorttitle{Rapid Gamma-ray Variability in PKS\,1510$-$089}
\shortauthors{Saito et al.}

\begin{document}

\title{Very Rapid High-Amplitude Gamma-ray Variability in Luminous Blazar PKS\,1510$-$089 Studied with Fermi-LAT}

\author{S.~Saito$^{1,\,2}$, \L .~Stawarz$^{1,\,3}$, Y.~T.~Tanaka$^4$, T.~Takahashi$^{1,\,2}$, G.~Madejski$^{5}$, and F.~D'Ammando$^{6}$}

\affil{$^1$ Institute of Space and Astronautical Science JAXA, 3-1-1 Yoshinodai, Chuo-ku, Sagamihara, Kanagawa 252-5210, Japan} 
\affil{$^2$ Department of Physics, Graduate School of Science, University of Tokyo, Hongo 7-3-1, Bunkyo, Tokyo 113-0033, Japan} 
\affil{$^3$ Astronomical Observatory, Jagiellonian University, ul. Orla 171, 30-244 Krak\'ow, Poland}
\affil{$^4$ Department of Physical Sciences, Hiroshima University, Higashi-Hiroshima, Hiroshima 739-8526, Japan}
\affil{$^5$ W.W. Hansen Experimental Physics Laboratory, Kavli Institute for Particle Astrophysics and Cosmology, Department of Physics and SLAC National Accelerator Laboratory, Stanford University, Stanford, CA 94305, USA.}
\affil{$^6$ Dip. di Fisica, Universit‡ degli Studi di Perugia and INFN, Via A. Pascoli, I-06123 Perugia, Italy}

\email{ssaitoh@astro.isas.jaxa.jp}

\label{firstpage}

\begin{abstract}

Here we report on the detailed analysis of the $\gamma$-ray light curve of a luminous blazar \pks\ observed in the GeV range with the Large Area Telescope (LAT) onboard the \textit{Fermi} satellite during the period 2011 September -- December. By investigating the properties of the detected three major flares with the shortest possible time binning allowed by the photon statistics, we find a variety of temporal characteristics and variability patterns. This includes a clearly asymmetric profile (with a faster flux rise and a slower decay) of the flare resolved on sub-daily timescales, a superposition of many short uncorrelated flaring events forming the apparently coherent longer-duration outburst, and a huge single isolated outburst unresolved down to the timescale of three-hours. In the latter case we estimate the corresponding $\gamma$-ray flux doubling timescale to be below one hour, which is extreme and never previously reported for any active galaxy in the GeV range. The other unique finding is that the total power released during the studied rapid and high-amplitude flares constitute the bulk of the power radiatively dissipated in the source, and a significant fraction of the total kinetic luminosity of the underlying relativistic outflow. Our analysis allows us to access directly the characteristic timescales involved in shaping the energy dissipation processes in the source, and to provide constraints on the location and the structure of the blazar emission zone in \pks.
\end{abstract}


\section{Introduction}
\label{sec:intro}

Blazars constitute a class of radio-loud active galactic nuclei for which the observed broad-band spectra are dominated by the beamed and highly variable emission of the innermost parts of relativistic jets \citep[see][]{Urry95}. Detailed and systematic investigation of the blazar variability at different wavelengths provides crucial information on the location and the structure of the energy dissipation zone(s) in blazar jets, as well as on the radiative and particle acceleration mechanisms involved in the production of the observed radiation. Recently, thanks to the operation of the modern high-energy instruments including ground-based Cherenkov telescopes and space-borne satellites, such studies can be performed also in the $\gamma$-ray regime. The observational results obtained in this way so far are rather striking. 

Nearby low-luminosity blazars of the BL Lac type were observed to be variable in the TeV range at the extremely short timescales of a few minutes \citep{HESS2155,MAGIC501}, challenging the homogeneous synchrotron-self-Compton emission models typically discussed in the context of the very high-energy (VHE; photon energies $\varepsilon_{\gamma} >0.1$\,TeV) $\gamma$-ray emission of BL Lacs \citep[see, e.g.,][]{Begelman08,Giannios09}. In high-luminosity blazars of the `Flat Spectrum Radio Quasar' type (FSRQs), equally spectacular high-amplitude flux changes on the sub-daily timescales in the GeV range have been established during the flaring states of several sources based on the data provided by the \textit{Fermi} Gamma-Ray Space Telescope (see \S\,4 below). These findings, especially when combined with the unexpected detections of a few FSRQs in the TeV range \citep{MAGIC279,MAGIC1222}, again put into question the validity of the one-zone inverse-Compton (IC) scenarios widely applied in the modeling of the high-energy (HE; $\varepsilon_{\gamma} \simeq 0.1-100$\,GeV) $\gamma$-ray continua of such objects \citep[e.g.,][]{Bottcher09,Tavecchio11}.

One of the FSRQs for which rapid HE outbursts have been seen with \textit{Fermi} is PKS 1510$-$089. This is a luminous blazar located at the redshift of $z=0.361$ (luminosity distance $d_{\rm L} \simeq 1.91$\,Gpc)\footnote{We assume $\Lambda$CDM cosmology with $\Omega_{\Lambda} = 0.73$, $\Omega_{\rm M} = 0.27$, and $H_{\rm 0} = 71$\,km\,s$^{-1}$\,Mpc$^{-1}$.}. The inner structure of the source is characterized by the dramatic morphological changes with the apparent superluminal velocities up to $\beta_{\rm app} \simeq 20$  \citep[e.g.,][]{Orienti11}, and the extreme mis-alignment between the milli-arcsec and arcsec-scale jets \citep[$\simeq 180\degr$ projected;][]{Homan02}, implying ultrarelativistic velocities of the emitting plasma (jet bulk Lorentz factor $\Gamma_j \simeq 20$) at small inclinations to the line of sight (jet viewing angle $\theta_j \simeq 3\degr$). The synchrotron emission of the inner jet in \pks\ peaks around the infrared frequencies \citep[e.g.,][]{Nalewajko12}; the UV segment of the source spectrum is contributed by accretion disk \citep[most clearly during the quiescent states;][]{DAmmando09}, and the X-ray continuum is dominated by the IC emission of the low-energy jet electrons \citep[e.g.,][]{Kataoka08}. \pks\ is also an established $\gamma$-ray emitter, detected in the HE range by the EGRET instrument on board \textit{CGRO}, \textit{Fermi}, and \textit{AGILE} satellites \citep{Hartman99,LAT1510,DAmmando11}, as well as in the VHE range by the \textit{H.E.S.S.} and \textit{MAGIC} telescopes \citep{HESS1510,MAGIC1510}.

Here we report on the detailed analysis of the spectacular $\gamma$-ray outbursts of PKS 1510$-$089 detected in the GeV range with LAT onboard the \textit{Fermi} satellite during the period 2011 September -- December (\S\,2). We study the profiles of the HE flares of \pks\ with different time binning, finding a variety of the temporal characteristics and variability patterns with the flux doubling timescales as short as below one hour (\S\,3). We discuss the energetics of the flares, concluding on the implication of the obtained results for the general understanding of the location and the structure of the energy dissipation zone in the relativistic jet of \pks\ and other luminous blazars of the FSRQ type (\S\,4).

\section{Data Analysis}

The \textit{Fermi}-LAT is a pair-conversion $\gamma$-ray telescope sensitive to photon energies from 20\,MeV up to 300\,GeV, characterized by the energy resolution typically $\sim10\%$ and the angular resolution ($68\%$ containment angle) better than $1\degr$ at energies above 1\,GeV. With its large field of view of 2.4\,sr it operates mostly in a survey mode scanning the entire sky every 3 hours \citep[for further details see][]{Atwood09}.

The data discussed in this paper were collected from MJD 55834 (2011 September 30) to MJD 55903 (2011 December 8), the period which overlapped with three major $\gamma$-ray outbursts of \pks. Only the events with energies greater than 100\,MeV and zenith angles $< 100\degr$ were included in this analysis. We selected good time intervals (GTIs) by using a logical filter of ``\texttt{DATA$\_$QUAL==1}'',  ``\texttt{LAT$\_$CONFIG==1}'', and ``\texttt{ABS(ROCK\_ANGLE)$<$52}''. The LAT data collected in that manner were analyzed with an unbinned likelihood analysis method using the standard analysis tool \texttt{gtlike}, which is a part of the \textit{Fermi}-LAT Science Tools software package (v9r27p1). The \texttt{P7SOURCE$\_$V6} set of instrument response functions was utilized. 

The $\gamma$-ray photons were extracted from a circular region of interest (ROI) with radius $10\degr$, centered at the radio position of the source. The Galactic diffuse emission template version ``\texttt{gal$\_$2yearp7v6$\_$v0.fits}'' and the isotropic diffuse emission template version ``\texttt{iso$\_$p7v6source.txt}'' were assumed in the modeling. The source model consisted of \pks\ and other point sources within the ROI and the surrounding $5\degr$-wide annulus taken from the second-year LAT catalog \citep[2FGL;][]{Nolan12}. Moreover, we included an additional variable $\gamma$-ray point source located at (R.A., Dec.)\,$=$\,(233.168, $-$13.311), i.e. about $6.4\degr$ away from \pks. This object, tentatively associated with the FSRQ TXS\,1530--131 \citep{Gasparrini11}, did not appear in the 2FGL, although it was clearly detected during the period considered here at relatively high flux level of $F_{>100\,{\rm MeV}} \simeq (1.1 \pm 0.1) \times 10^{-7}$\,ph\,cm$^{-2}$\,s$^{-1}$.

In the following analysis of the source light curves, we fixed the
fluxes of the diffuse emission components at the values obtained by
fitting the data collected over the entire period discussed
here. Normalizations of all the point sources within $10\degr$ radius
around \pks\ were set free, while the other relevant parameters of the
neighboring $\gamma$-ray emitters were fixed following the second-year
LAT catalog. For each time bin analyzed, point sources with the test
statistic (TS) values $\leq 0$ \citep[see][]{Mattox96} were removed
from the source model. We approximated the $\gamma$-ray
continuum of \pks\ with a simple power-law model, keeping 
photon indices free during the likelihood analysis of the source 
spectra for all the time binnings considered (down to 3\,h).
The power-law fits were acceptable in each case, indicating that more 
complex spectral shapes, with more degrees of freedom, were formally 
not required \citep[but see][]{Orienti12}.

\section{Results}

\begin{figure}[h]
\begin{center}
\includegraphics[width=\columnwidth, bb=0 0 1000 500,clip]{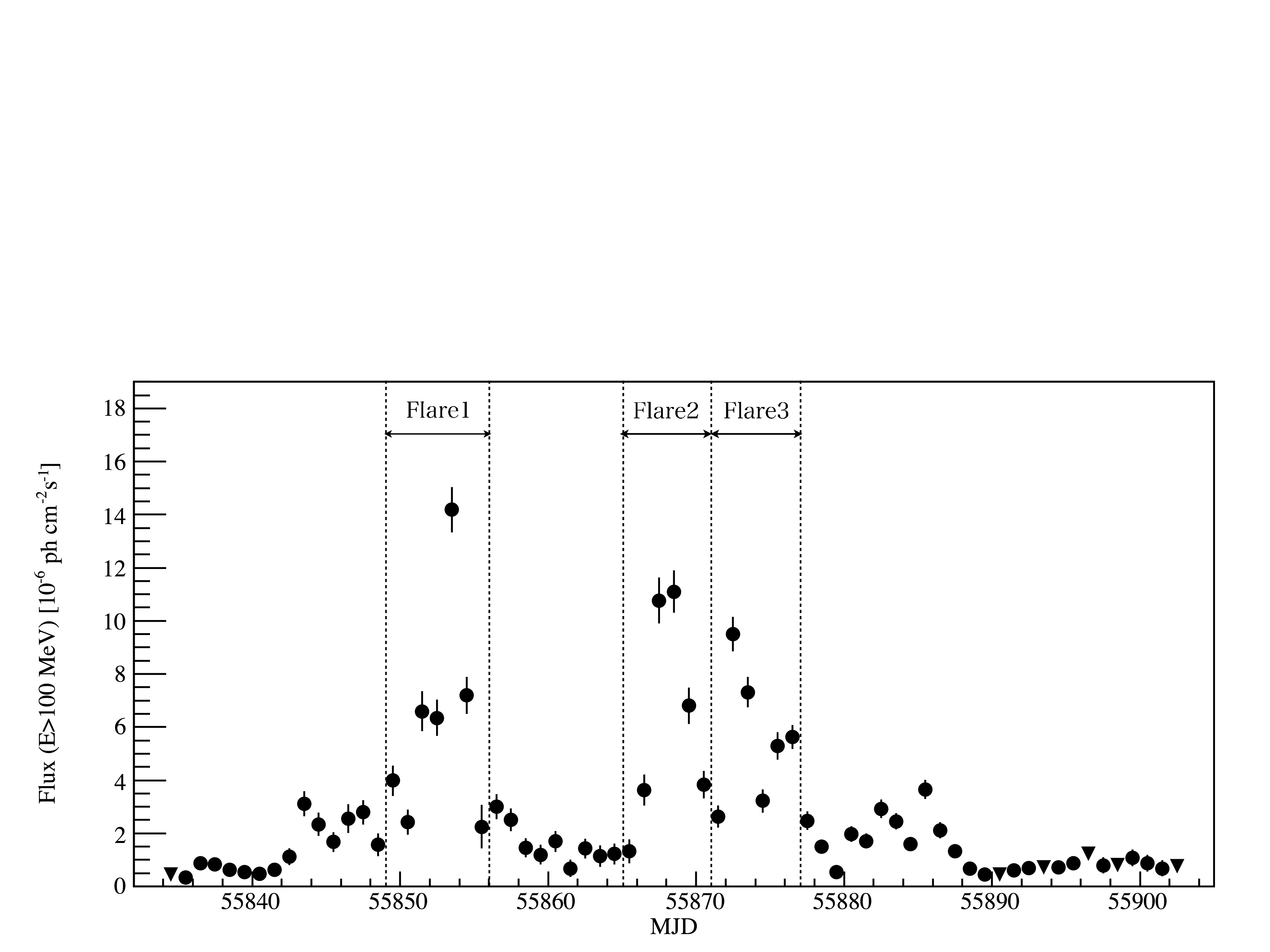}
\caption[]{Daily $\gamma$-ray light curve of \pks\ during the period MJD 55834--55903 analyzed in this paper. $95\%$ flux upper limits are represented by triangles. Horizontal lines separating the three major flares are chosen arbitrarily just to guide the eye. \label{lc}}
\end{center}
\end{figure}

Figure\,\ref{lc} presents the daily $\gamma$-ray light curve of \pks\ at
photon energies 0.1--300\,GeV during the period analyzed in this paper. In our analysis, $95\%$ confidence level flux upper limits correspond to the detection significance values TS $<$ 10, which is a conventional choice in the analysis of daily-binned light curves of bright LAT sources \citep[see, e.g.,][]{Tavecchio10,Orienti12}. As shown, during the discussed time interval three major high-amplitude $\gamma$-ray outbursts of the source were detected with photon fluxes $F_{>100\,{\rm MeV}} \gtrsim 10^{-5}$\,ph\,cm$^{-2}$\,s$^{-1}$ (see Table\,\ref{maxima}), and flux doubling timescales less than a day. The excellent photon statistics allowed us to study these flares with shorter time binning, down to the minimum 3\,h dictated by the survey mode of the LAT instrument.

\begin{figure}[h]
\begin{center}
\includegraphics[width=\columnwidth,bb=0 0 600 650,clip]{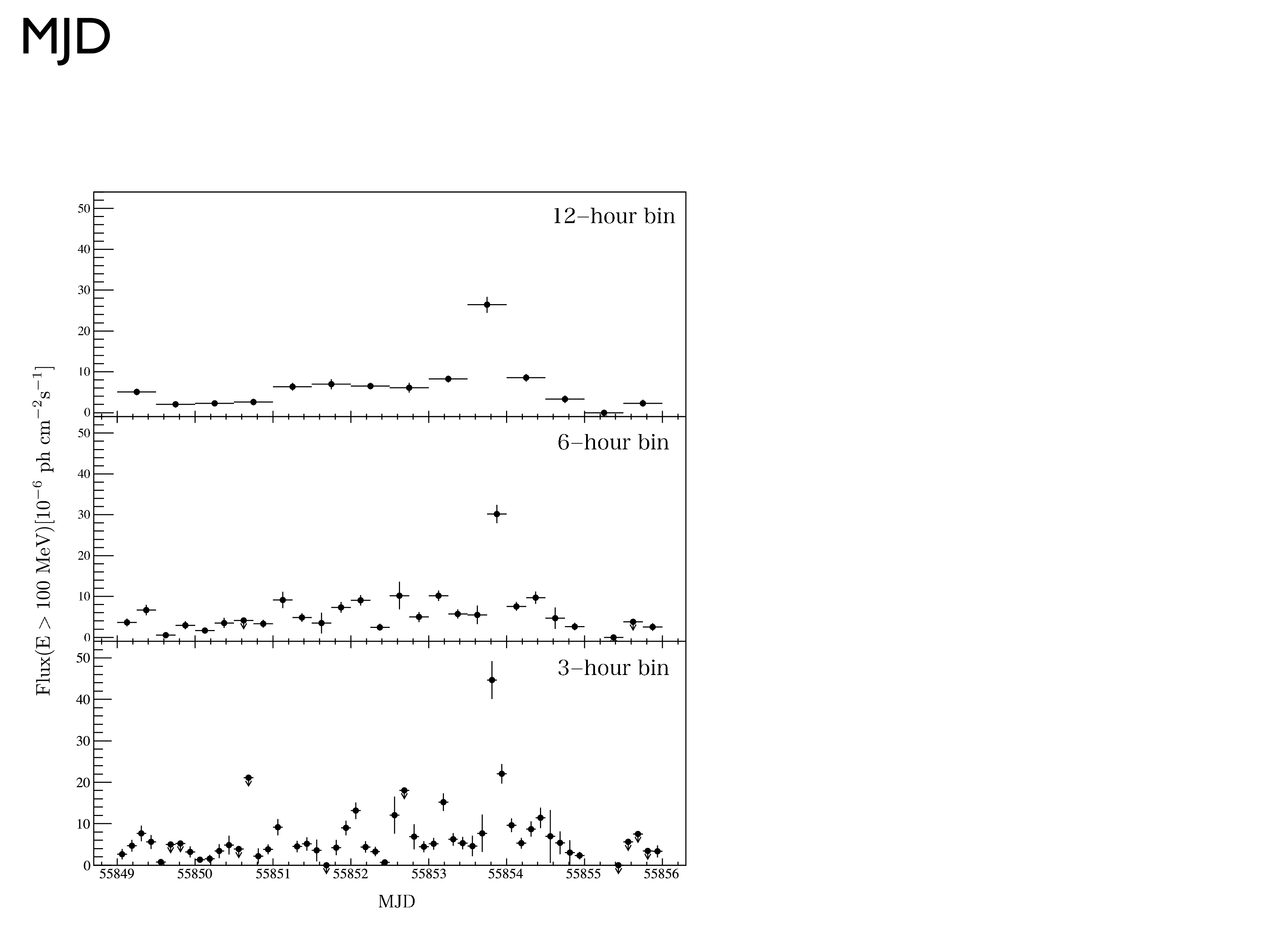}
\caption[]{\textit{Fermi}-LAT light curves of \pks\ around the time of the first major $\gamma$-ray outburst, binned in the intervals of 12\,h, 6\,h, and 3\,h (upper, middle, and lower panels, respectively). \label{f1}}
\end{center}
\end{figure}

Figure\,\ref{f1} presents the light curves of \pks\ around the time of the first major $\gamma$-ray outburst, binned in the intervals of 12\,h, 6\,h, and 3\,h (upper, middle, and lower panels, respectively). As shown, the rising segment of the flare is unresolved down to the timescale of 3\,h, and this is a truly unique finding for a luminous blazar. Previously all the high-amplitude flux changes of FSRQs detected in the GeV range were characterized by longer ($\geq 1$\,d) timescales, while any shorter variability consisted of a small-amplitude flickering only (see \S\,4 below). Here, instead, the recorded flux increases from about $F_1 \simeq 7 \times 10^{-6}$\,ph\,cm$^{-2}$\,s$^{-1}$ up to $F_2 \simeq 45 \times 10^{-6}$\,ph\,cm$^{-2}$\,s$^{-1}$ within $\Delta t = 3$\,h, giving formally the flux doubling timescale of $\tau_{\rm d} = \Delta t \times \ln 2 / \ln (F_2/F_1) \simeq 1$\,h only, or equivalently the exponential growth timescale $\tau_{\rm d} / \ln 2 \simeq 1.5$\,h. This value should be considered as an upper limit only, because of a limited exposure of \pks\ during the analyzed 3\,h visibility window. Interestingly, the decay segment of the flare seems to be marginally resolved with the 3\,h binning, implying the $e$-folding decay timescale of about 4\,h. The evaluated photon index $\Gamma_{\gamma} \simeq 2.0$, together with the large Compton dominance established for \pks\ \citep[e.g.,][]{Nalewajko12}, imply that the bulk of the radiative energy released during the flare is contained within the HE range.

\begin{figure}[h]
\begin{center}
\includegraphics[width=\columnwidth, bb=0 0 600 650,clip]{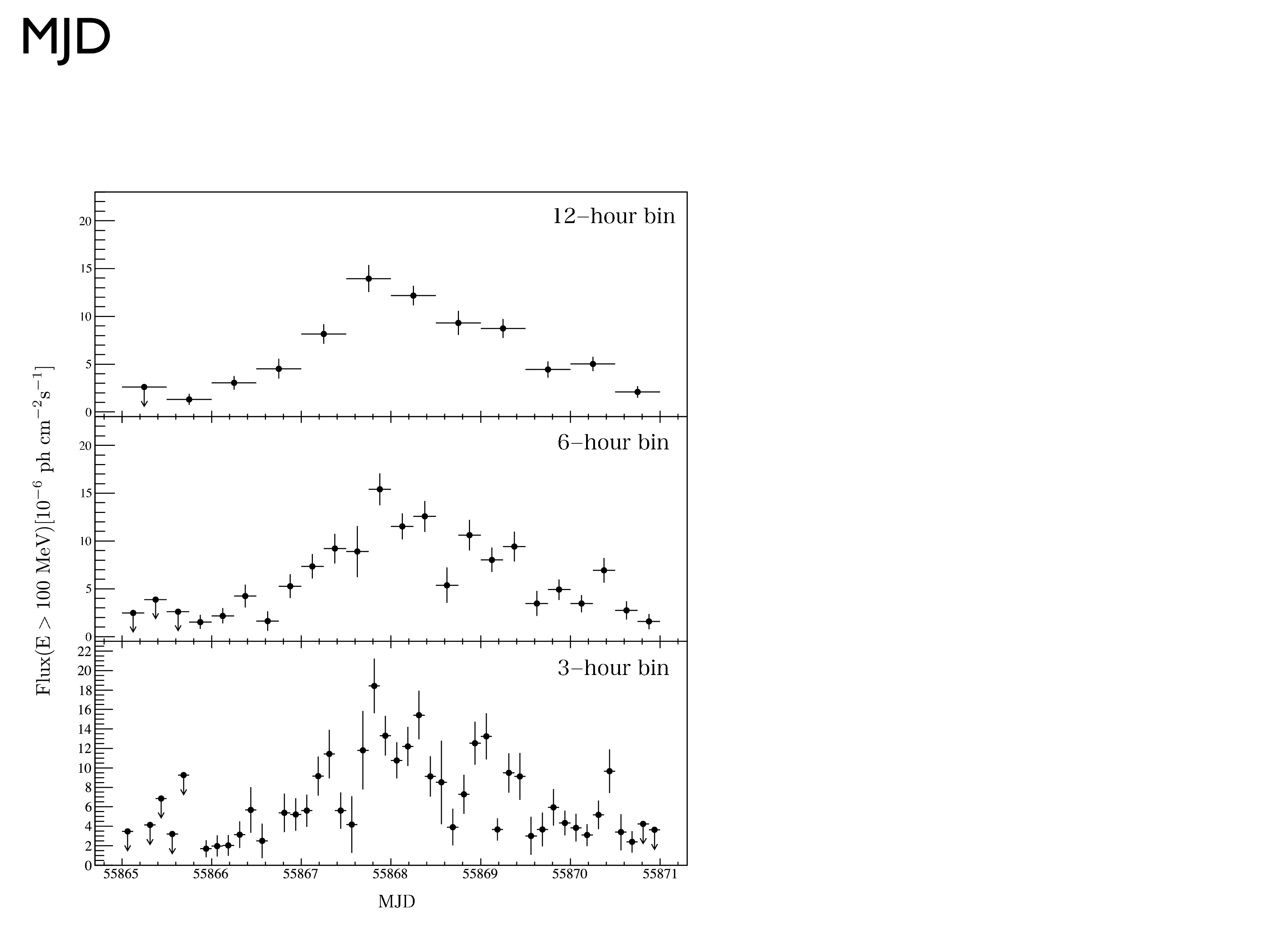}
\caption[]{Same as in Figure\,\ref{f1} but for the second major $\gamma$-ray outburst. \label{f2}}
\end{center}
\end{figure}

The second $\gamma$-ray outburst of \pks\, for which the LAT light curves in 12, 6, and 3\,h bins are presented in Figure\,\ref{f2}, constitutes a very different case. Here the flare seems to be resolved with 12\,h-binning, displaying shorter exponential growth and a slower linear decay, as expected in most of the models of FSRQs' variability involving a fast injection of accelerated electrons and their slower radiative cooling dominated by the Comptonization of the soft photons produced externally to the jet \citep[e.g.,][]{Sikora01}. However, with the minimum 3\,h binning a significant sub-structure of a flare becomes prominent, consisting of several apparently chaotic and unresolved yet still large-amplitude events, often characterized by the flux doubling timescales $< 3$\,h. This clearly illustrates the fact that with the limited time resolution, the apparent profiles of high-energy outbursts in blazar sources may not reflect the exact temporal characteristics of the source flux changes.

\begin{figure}[h]
\begin{center}
\includegraphics[width=\columnwidth, bb=0 0 600 650,clip]{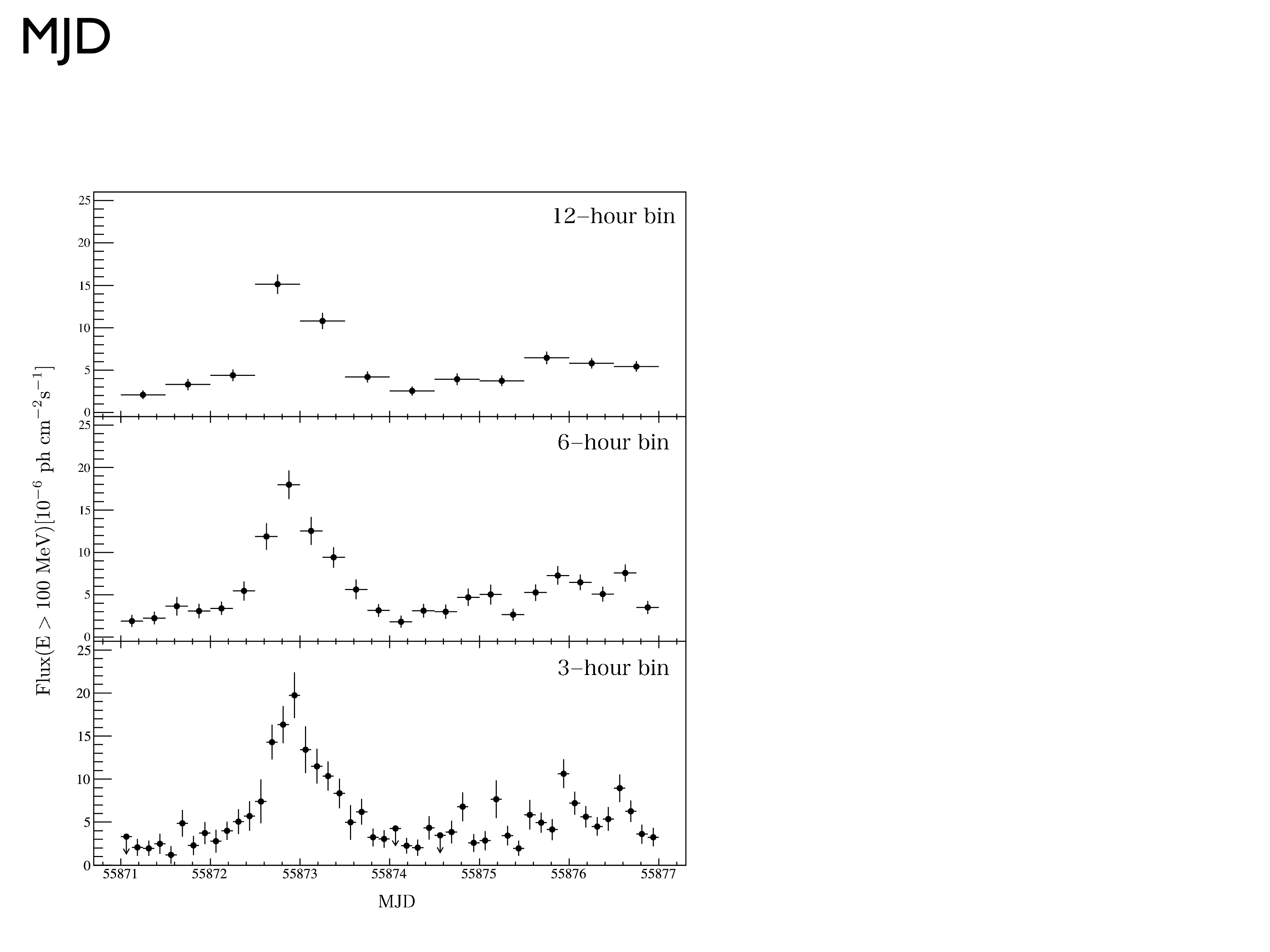}
\caption[]{Same as in Figure\,\ref{f1} but for the third major $\gamma$-ray outburst. \label{f3}}
\end{center}
\end{figure}

Finally, Figure\,\ref{f3} presents the light curves of \pks\ around the time of the third major $\gamma$-ray outburst, binned again in the intervals of 12\,h, 6\,h, and 3\,h. As shown, the flare seems to be nicely resolved in short binning, displaying a moderately asymmetric profile with a faster flux increase (doubling timescale between 3\,h and 6\,h), and a longer flux decay ($e$-folding timescale of about 11\,h). However, we cannot exclude a possibility that with even shorter binning of the light curve, this smooth and seemingly coherent flaring event would be decomposed into a series of rapid overlapping but not necessarily related sub-events.

\section{Discussion and Conclusions}

Strong HE flares from \pks\ have been detected with \textit{AGILE} in 2008 March and 2009 March \citep{DAmmando09,DAmmando11}, with the daily-integrated peak fluxes of $F_{>100\,{\rm MeV}} \simeq (3-7) \times 10^{-6}$\,ph\,cm$^{-2}$\,s$^{-1}$, photon indices $\Gamma_{\gamma} \simeq 2.0$, and the flux doubling timescales of the order of a day. Similar results were reported by the \textit{Fermi}-LAT Collaboration for the period 2008 September -- 2009 June \citep{LAT1510}, consisting of the detection of several outbursts lasting for a few/several days with the daily-integrated peak fluxes $F_{>100\,{\rm MeV}} \simeq (2-8) \times 10^{-6}$\,ph\,cm$^{-2}$\,s$^{-1}$, photon indices $\Gamma_{\gamma} \lesssim 2.5$, and the characteristic $e$-folding flux variability timescale estimated to be of the order of 3\,h (for the minimum 6\,h-binning of the light curve applied). 

The 2009 March and April \textit{Fermi}-LAT data for \pks\ were analyzed also by \citet{Tavecchio10}, who found significant flux changes by a factor of two or more occurring on the timescale of 6\,h, with approximately symmetrical flare profiles, and also well-defined events characterized by even shorter variability, which however could not be claimed at a high confidence level due to large errors related with lower flux level of the source. Similar behavior with the flux doubling timescales of the order of several hours has been reported for some other flaring FSRQs observed with LAT, namely PKS\,1454$-$354 \citep{LAT1454}, PKS\,1502+106 \citep{LAT1502}, 3C\,273 \citep{LAT273}, and 3C\,454.3 \citep{LAT454,Tavecchio10}. Finally, \citet{Foschini11} set the upper limits of 2\,h on the observed $\gamma$-ray doubling time scale in quasars 3C\,454.3, 3C\,273, and PKS\,1222+216 during their bright flaring states with daily photon fluxes $F_{>100\,{\rm MeV}}$ exceeding $10^{-5}$\,ph\,cm$^{-2}$\,s$^{-1}$.

The isotropic \emph{daily-averaged} HE luminosity of the first flare analyzed here is $L_{\rm \gamma,\, iso} \simeq 7 \times 10^{48}$\,erg\,s$^{-1}$. The corresponding total power emitted in $\gamma$-rays \citep[i.e., the power as would be measured by the detector completely surrounding the emitting region; e.g.,][]{Sikora97} is therefore $L_{\rm \gamma,\, em} \simeq L_{\rm \gamma,\, iso} / 4 \Gamma_{\rm j}^2 \simeq 5 \times 10^{45}$\,erg\,s$^{-1}$ (assuming $\Gamma_{\rm j} \simeq 20$; see \S\,1), which is almost exactly the same as the total kinetic power of the \pks\ jet emerging from broad-band modeling based on different datasets and model assumptions, $L_{\rm j} \gtrsim 5 \times 10^{45}$\,erg\,s$^{-1}$, and also as the observed UV disk luminosity in the system, $L_{\rm disc} \simeq 5 \times 10^{45}$\,erg\,s$^{-1}$ \citep{Kataoka08,DAmmando09,LAT1510}. This implies that, during the discussed flaring event, the power dissipated in the jet within less than a day and emitted as $\gamma$-ray photons constitutes the bulk of the total kinetic luminosity carried out by the outflow, $L_{\rm \gamma,\, em} /L_{\rm j} \lesssim 1$, and also a substantial  fraction of the entire available accretion power, $L_{\rm \gamma,\, em} / L_{\rm acc} \simeq 0.1$ (assuming the standard $\eta_{\rm disk} \simeq 10\%$ radiative efficiency for the accretion disk, $L_{\rm acc} \simeq L_{\rm disc} /  \eta_{\rm disk} \simeq 5 \times 10^{46}$\,erg\,s$^{-1}$). Note in this context that, for the black hole mass in the system $\mathcal{M}_{\rm BH} \simeq 5 \times 10^8 M_{\odot}$ \citep{LAT1510}, the active nucleus in \pks\ accretes at the maximum Eddington rate, $L_{\rm acc} \sim L_{\rm Edd}$. A very similar set of the source parameters, implying the extremely efficient conversion of the accretion power to the jet $\gamma$-ray luminosity has been established before by \citet{Tanaka11} for the analogous blazar PKS\,1222+216 observed with LAT during its flaring state.

Equally interesting is the analysis of the emerging timescales and the related (via the causality arguments) emission zone spatial scales. In particular, the observed flux doubling timescale of $\tau_{\rm d} \simeq 1$\,h and the bulk Lorentz factor $\Gamma_{\rm j} \simeq 20$ (equal by assumption to the jet Doppler factor, consistently with the expected jet inclination $\theta_{\rm j} \simeq 3\degr$) give the spatial scale of the emitting region $R_{\rm var} \leq c \tau_{\rm d} \Gamma_{\rm j} / (1+z) \simeq 1.5 \times 10^{15}$\,cm. Meanwhile, the gravitational radius of the \pks\ supermassive black hole is $r_{\rm g} = G \mathcal{M}_{\rm BH} / c^2 \simeq 7 \times 10^{13}$\,cm. Assuming a very basic scenario in which the scale of the event horizon sets a lower limit on the spatial scale of the jet disturbances that can be identified with the zones of the enhanced energy dissipation, one should expect such structures, when created near the black hole, to be advected along the outflow and to release the bulk of their power around $r_{\rm em} \simeq \Gamma_{\rm j}^2 r_{\rm g} \simeq 3 \times 10^{16}$\,cm distances from the core \citep{Begelman08}. The characteristic radial scale of the outflow at that point is then expected to be approximately $R_{\rm j} \simeq r_{\rm em} / \Gamma_{\rm j} \simeq 1.5 \times 10^{15}$\,cm, following the standard expectation for the jet opening angle $\simeq 1/\Gamma_{\rm j}$. The agreement between the derived values of  $R_{\rm var}$ and $R_{\rm j}$ is striking.

This identified blazar zone would be located inside the region of the highest ionization of the broad line-emitting circumnuclear clouds (`broad line region'; BLR), for which the characteristic scale in the discussed system is $r_{\rm BLR} \simeq 2 \times 10^{17}$\,cm \citep{LAT1510,Nalewajko12}. There the energy density provided by the line-emitting clouds should exceed energy densities of the other photon fields in the jet rest frame, and hence the dominant production of the $\gamma$-ray photons should be related to the IC upscattering of the UV emission (observed energies $\varepsilon_0 \simeq 10$\,eV) reprocessed with the $\xi_{\rm BLR} = 10\%$ efficiency within the BLR \citep[e.g.,][]{Ghisellini09,Sikora09}. The corresponding cooling timescale for the electrons emitting $\gamma$-rays with the energies of $\varepsilon_{\gamma} = 100$\,MeV, as measured in the observer frame, would then be $\tau_{\rm rad} \simeq (3 m_e c/4 \sigma_T u'_{\rm BLR}) \times [\varepsilon_0 (1+z)/\varepsilon_{\gamma}]^{1/2} \gtrsim 10$\,min, for the jet comoving BLR photon energy density $u'_{\rm BLR} \simeq \xi_{\rm BLR} \, L_{\rm disk} \Gamma_{\rm j}^2 / 4 \pi r_{\rm BLR}^2 c \simeq 10$\,erg\,cm$^{-3}$. This timescale is shorter by a factor of $10-50$ than the observed $e$-folding decay timescales of the flares, implying that the observed flux decrease is shaped not solely by the radiative energy losses, but instead by a combination of different factors. These other factors may be related either to the geometry and sub-structure of the emitting region \citep[e.g.,][]{Tanihata01}, or to a residual particle acceleration still ongoing after the peak of a flare.

The `near-dissipation zone' scenario, with the dominant emission region located relatively close to the central engine ($\lesssim 0.1$\,pc), was advocated in the literature for FSRQs in general based on the modeling of their HE $\gamma$-ray spectra \citep{Poutanen10}. The complication arises, however, due to the aforementioned detection of a few FSRQs, including \pks\ and PKS\,1222+216, at TeV photon energies \citep[see][]{Tanaka11,Tavecchio11}. The emerging agreement is that such VHE emission, if detected, must be produced instead at further distances from the core, i.e. beyond the characteristic scale of the circumnuclear dust ($> 0.1$\,pc).

The broad-band variability studies enabled by the
multiwavelength campaigns carried out during the recent years revealed
rather complex behavior of \pks\, with no HE/X-ray correlations, weak
HE/UV correlations, and pronounced HE/optical correlations;
some of the HE/optical correlations were found to consist of
optical flares lagging the HE outbursts by several days, while
other events showed a time-lag consistent with zero
\citep{Marscher10,LAT1510,DAmmando11}. In addition, the
apparent correlations between some major $\gamma$-ray flares with the
coherent rotations of the optical polarization vector (swings by
$>360\degr$), as well as with the structural changes in the inner radio
jet, have been discussed by \citet{Marscher10} and
\citet{Orienti12} as important findings supporting the `far-dissipation zone' scenario for the source. It is not clear, however, if this model can be easily reconciled with the short variability timescales found in our analysis for the powerful $\gamma$-ray outbursts dissipating the bulk of the total kinetic luminosity of the jet in the system. A substantial sub-structure of the outflow, consisting of a `jets-in-jet' configuration \citep{Giannios09}, or of a highly turbulent relativistic plasma passing through a standing shock \citep[in which case one could expect an enhanced energy dissipation only in a small fraction of the jet; see][]{Marscher12}, may be possible solutions.\footnote{During the review process of this paper, a comprehensive analysis by \citet{Brown13} of the overlapping dataset, with similar results and conclusions to our work, was posted on the arXiv.}
 
\section*{Acknowledgments}

S.~S. receives financial support from JSPS. \L .~S. was supported by Polish NSC grant DEC-2012/04/A/ST9/00083. 

The \textit{Fermi}-LAT Collaboration acknowledges support from a number of agencies and institutes for both development and the operation of the LAT as well as scientific data analysis. These include NASA and DOE in the United States, CEA/Irfu and IN2P3/CNRS in France, ASI and INFN in Italy, MEXT, KEK, and JAXA in Japan, and the K.~A.~Wallenberg Foundation, the Swedish Research Council and the National Space Board in Sweden. Additional support from INAF in Italy and CNES in France for science analysis during the operations phase is also gratefully acknowledged.

{}

\newpage

\begin{table}[h]
{\footnotesize
\noindent
{\caption[] {\label{maxima} Major $\gamma$-ray flares of \pks.}}
\begin{center}
\begin{tabular}{ccc}
\hline\hline
MJD & $F_{>100\,{\rm MeV}}$ & $\Gamma_{\gamma}$\\
(1) & (2) & (3)\\
\hline
55853.5--55854.5 & $14.86 \pm 0.89$ & $1.97 \pm 0.04$\\
55867-55869 & $10.95 \pm 0.57$ & $2.21\pm 0.04$\\
55872--55874 & $8.39 \pm 0.44$ & $2.19\pm 0.04$\\
\hline\hline
\end{tabular}
\end{center}
(1) Dates of the three major $\gamma$-ray flux maxima in the daily-binned light curve of \pks; (2) photon fluxes measured at the flux maxima in the units of [$10^{-6}$\,ph\,cm$^{-2}$\,s$^{-1}$], averaged over the specified time intervals; (3) the corresponding photon indices.
}
\end{table}

\newpage

\end{document}